**Deposition and growth of the AlCoCuFeNi high entropy alloy thin film: molecular dynamics simulation**

**O. I. Kushnerov, V. F. Bashev, and S.I. Ryabtsev**

O. I. Kushnerov(⌧)·V. F. Bashev ·S.I. Ryabtsev

Oles Honchar Dnipro National University, Gagarin ave., 72, Dnipro, 49010, Ukraine

e-mail: kushnrv@gmail.com

**Abstract** The growth of a thin film of a high-entropy AlCoCuFeNi alloy on a substrate of silicon (100) was studied using molecular dynamics modeling. The simulation was carried out using the embedded atom model to describe the interaction among Al–Co–Cu–Ni–Fe. Interaction between the atoms of Al, Co, Cu, Fe, Ni, and the Si substrate was modeled using the Lennard–Jones potential, and the interaction between the silicon atoms was described using the Stillinger-Weber potential. Total simulation time has reached 50 ns. It was found that at the first stage of deposition small clusters were formed and the process of crystallization started after ~ 5 ns of simulation, at the characteristic sizes of clusters of about 2 nm. At the end of the simulation, after the 50 ns of modeling, the simulated film contains a face-centered cubic phase, a body-centered cubic phase, a hexagonal close-packed phase, and an amorphous phase. An analysis of the radial distribution of atoms made it possible to determine the distances between the nearest neighbors and estimate the lattice parameters of these phases.

## 1. Introduction

High-entropy alloys (HEAs) are a relatively new class of metallic materials developed by Yeh et al. in 2004. Such alloys usually contain from 5 to 13 basic elements in equiatomic or close to equiatomic concentrations (from 5 to 35%). Due to the high mixing entropy, multicomponent high entropy alloys typically consist of simple solid solutions with face-centered cubic (FCC) or body-centered cubic (BCC) lattices. Many high-entropy alloys possess unique properties, such as wear-resistance, resistance to corrosion and oxidation, radiation resistance, high hardness, and strength [1–5]. Thus, the high-entropy alloys may find use as materials for nuclear reactors applications, medicine, electronics devices, mechanical equipment, rocket casings, and engines, etc. The majority of HEAs were investigated in the as-quenched or homogenized state, whereas much less attention was paid to the study of thin films of high-entropy alloys. However, due to the very high cooling rates required to avoid the formation of complex crystallized phases and freeze the low-ordered structures, HEA is sometimes difficult to synthesize as bulk materials. The study of the growth characteristics of complex multi-element HEA films is an important topic of research considering that thin films can be used in various fields of technology [6]. Several papers [7-9] have been devoted to the study of thin film growth processes of one of the earliest and best-studied HEA of the AlCoCrCuFeNi system. However, a five-element AlCoCuFeNi HEA was later developed. This alloy does not contain Cr, which is commonly used in HEAs, and helps to improve their performance, but significantly increases the cost of the alloy. In this work, the processes of deposition and growth of thin films of AlCoCuFeNi HEA on the surface of a monocrystalline silicon substrate were investigated by classical molecular dynamics (MD) simulation.

## 2. Simulation model and methods

Classical MD simulation studies of deposition and growth of thin films of AlCoCuFeNi HEA have been performed using Large Scale Atomic/Molecular Massively Parallel Simulator (LAMMPS, Sandia National Laboratory, USA) [10]. Visualization of snapshots and the coordination analysis were done using open visualization tool software OVITO [11]. A three-dimensional cell with periodic boundary conditions only in two horizontal directions x and y was used. The free boundary condition was used in the vertical *z*-direction to ensure surface growth. The monocrystalline Si (100) substrate was 10×10×2.2 nm.

To prevent substrate displacement due to the impact of adatoms on the upper surface of the substrate, the positions of several lower atomic planes were fixed. The intermediate region of the substrate above the stationary atomic planes was controlled by a Berendsen thermostat and maintained at a temperature of 300 K which is similar to the experimental temperature mode to ensure isothermal growth conditions. The atoms of the upper layers of the substrate were not subjected to thermostating and were able to move freely under the impact of the deposited atoms (Fig.1). The 5 atoms of metals Al, Co, Cu, Fe, and Ni in the equimolar ratio were directed to a substrate every 10 ps, this leads to the atomic flux density of $5 \cdot 10^{23}$ atom/($cm^2 \cdot s$). The rate of 0.5 atom/ps was carefully chosen to allow sufficient time for thermal relaxation with a Berendsen thermostat (i.e. energy relaxation of the substrate after collision). It should be borne in mind that the atomic flux in the MD simulation is much higher than in the experiment, but the obtained data on morphology, structure, and adhesion coefficient are quite adequate [9]. The velocities of the deposited atoms corresponded to the mean kinetic energy of 0.1 eV. The initial positions of the deposited atoms were chosen randomly in the $\{x, y\}$ plane above substrate so that they were outside the range of the potential that describes the interaction in the *z*-direction. All of Al, Co, Cu, Fe, and Ni atoms were deposited from an initial position 15.7–16 nm above the substrate.

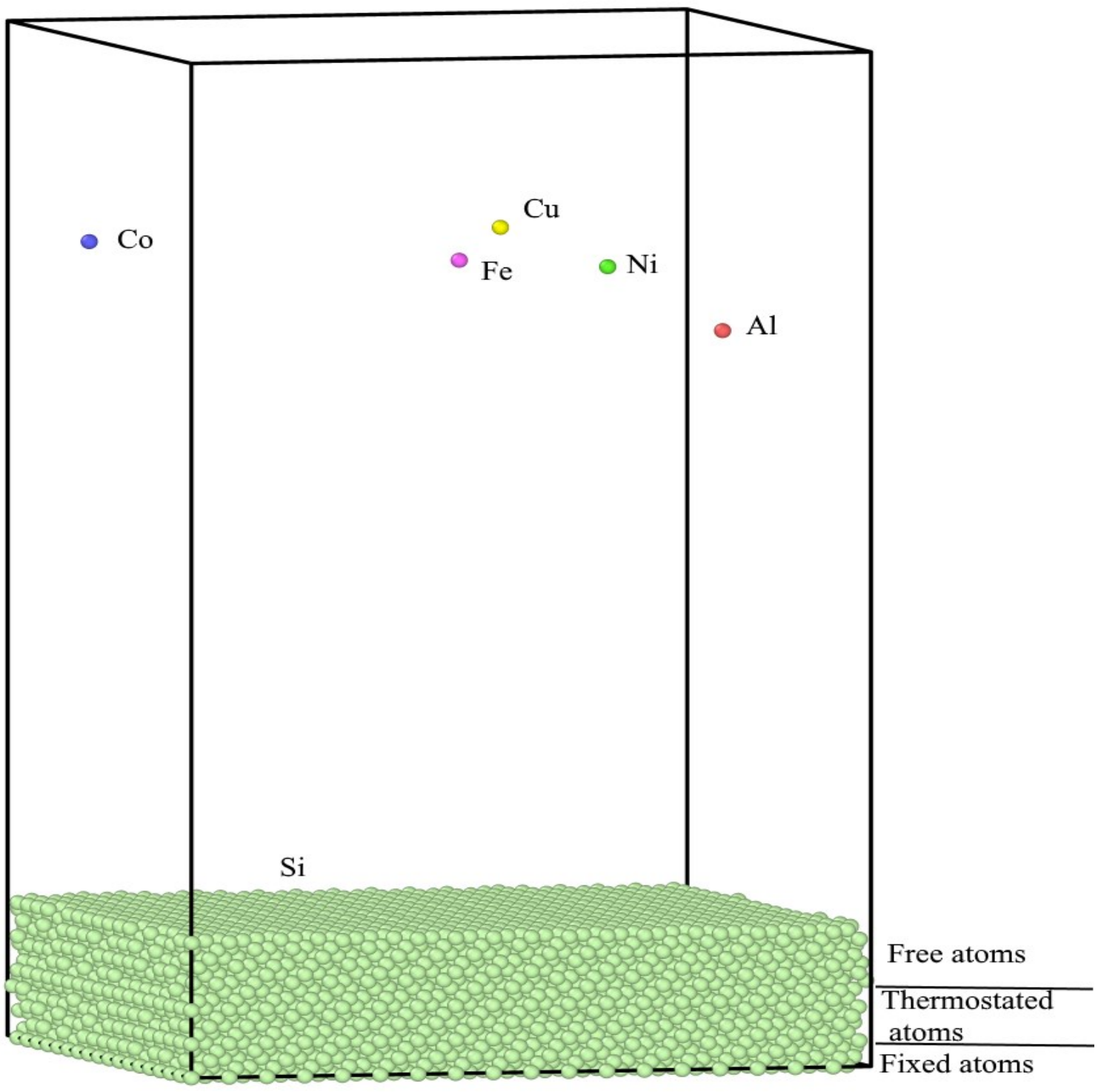

**Fig. 1** Schematic picture of the simulation model

The simulation was carried out using the embedded atom model (EAM) [12] to describe the interaction among Al–Co–Cu–Ni–Fe. EAM potentials include empirical formulas that simulate material characteristics such as surface energies, lattice constants, and the heat of solution. The EAM potential is based on the ideas of density functional theory, which generally state that the energy of a solid is a unique function of electron density.

The EAM potential is expressed as a multi-body potential energy function in the following form [12]

$$E = \sum_i f_i(\bar{\rho}_i) + \frac{1}{2}\sum_i \sum_{j \neq i} \phi_{ij}(r_{ij}) \tag{1}$$

where $f_i(\bar{\rho}_i)$ is the embedding energy function, $\phi_{ij}(r_{ij})$ is a pair interaction potential energy as a function of the distance $r_{ij}$ between atoms i and j that have chemical sorts *a* and *b*, $\bar{\rho}_i$ is the local host electron density. The energy *E* is the total energy of the atomistic system which comprises the sum of the embedding energy (first term of Eq.(1)) and the short-range pair potential energy (the second term). In the present research, the EAM potential developed by Zhou et al. [13] was used.

The interaction between the atoms of Al, Co, Cu, Fe, Ni, and the Si substrate was modeled using the Lennard–Jones potential

$$E_{ij}(r_{ij}) = 4\varepsilon\left[\left(\frac{\sigma}{r_{ij}}\right)^{12} - \left(\frac{\sigma}{r_{ij}}\right)^{6}\right] \tag{2}$$

with the parameters summarized in [7,8 ], and the interaction between the silicon atoms was described using the Stillinger-Weber potential [14].

## 3. Results and Discussion

During simulation 25000 atoms were deposited to the substrate, which corresponds to the nominal equiatomic AlCoCuFeNi composition, but some of them escaping the surface and the resulting film contained only 24433 atoms. So the sticking coefficient was ≈ 0.98. The composition of the simulated HEA film and the target composition is shown in Tab.1.

**Table 1** The element composition in a simulated thin film of AlCoCuFeNi HEA

| Number of atoms | Al | Co | Cu | Fe | Ni | Total |
|---|---|---|---|---|---|---|
| nominal composition | 5000 | 5000 | 5000 | 5000 | 5000 | 25000 |
| resulting film | 4909 | 4877 | 4865 | 4889 | 4893 | 24433 |

Thus, the final composition of the simulated AlCoCuFeNi film was quite close to equiatomic.

According to the results of the simulation, it was found that in the initial stages of growth (0 – 2 ns) the film represents an islet-like structure formed by small clusters (Fig.2). Later, the growth of the islands and the formation of continuous coating has begun. Due to the limited number of deposited atoms, it was impossible to obtain a completely continuous film even after 50 ns of simulation time. The film obtained at the end of the simulation is shown in Fig.3. The maximum thickness of the film has reached 4.3 nm.

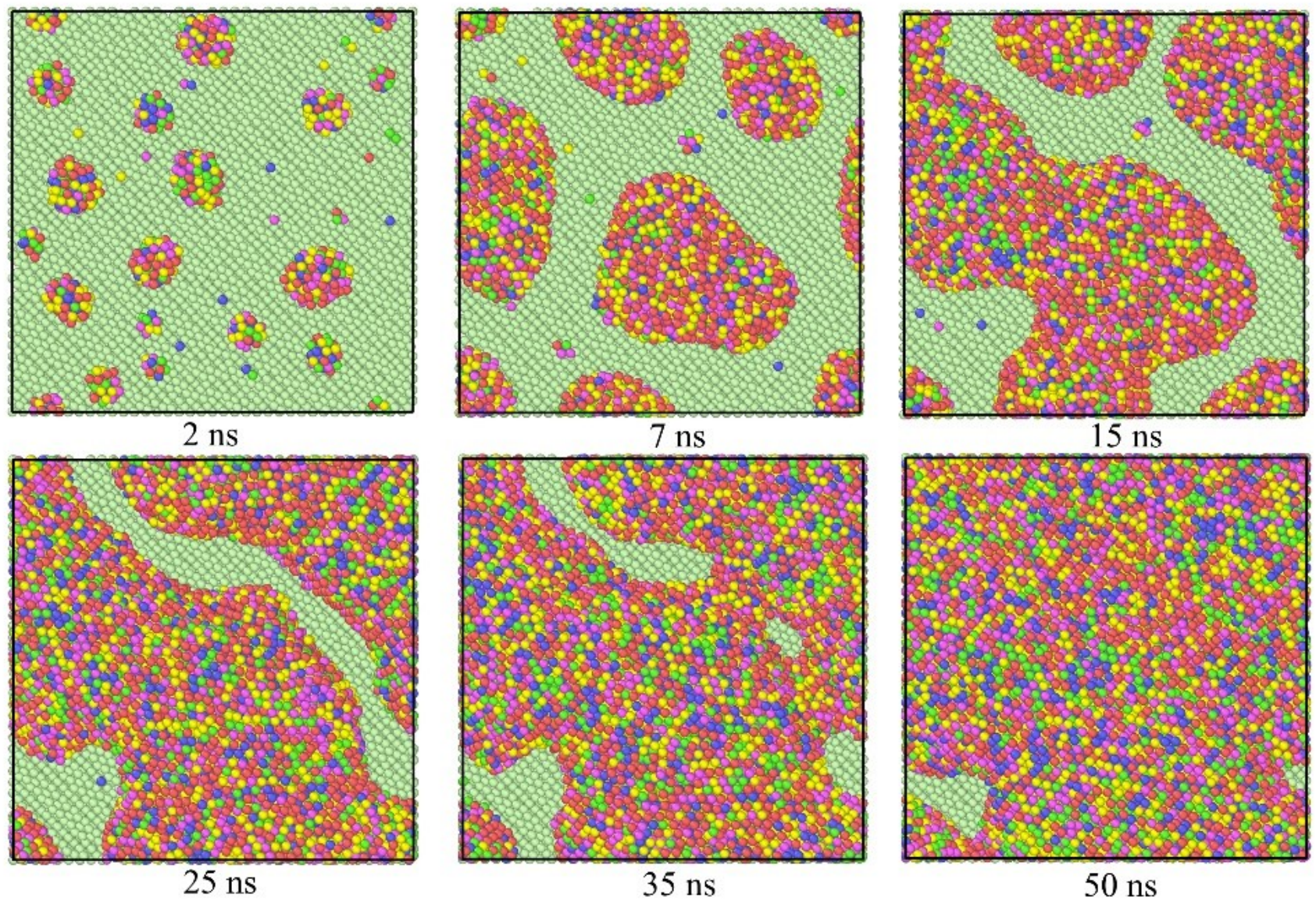

**Fig. 2** Snapshots of AlCoCuFeNi thin film deposited on Si substrate at different times (top view): ● – Al, ● – Co, ● – Cu, ● – Fe, ● – Ni.

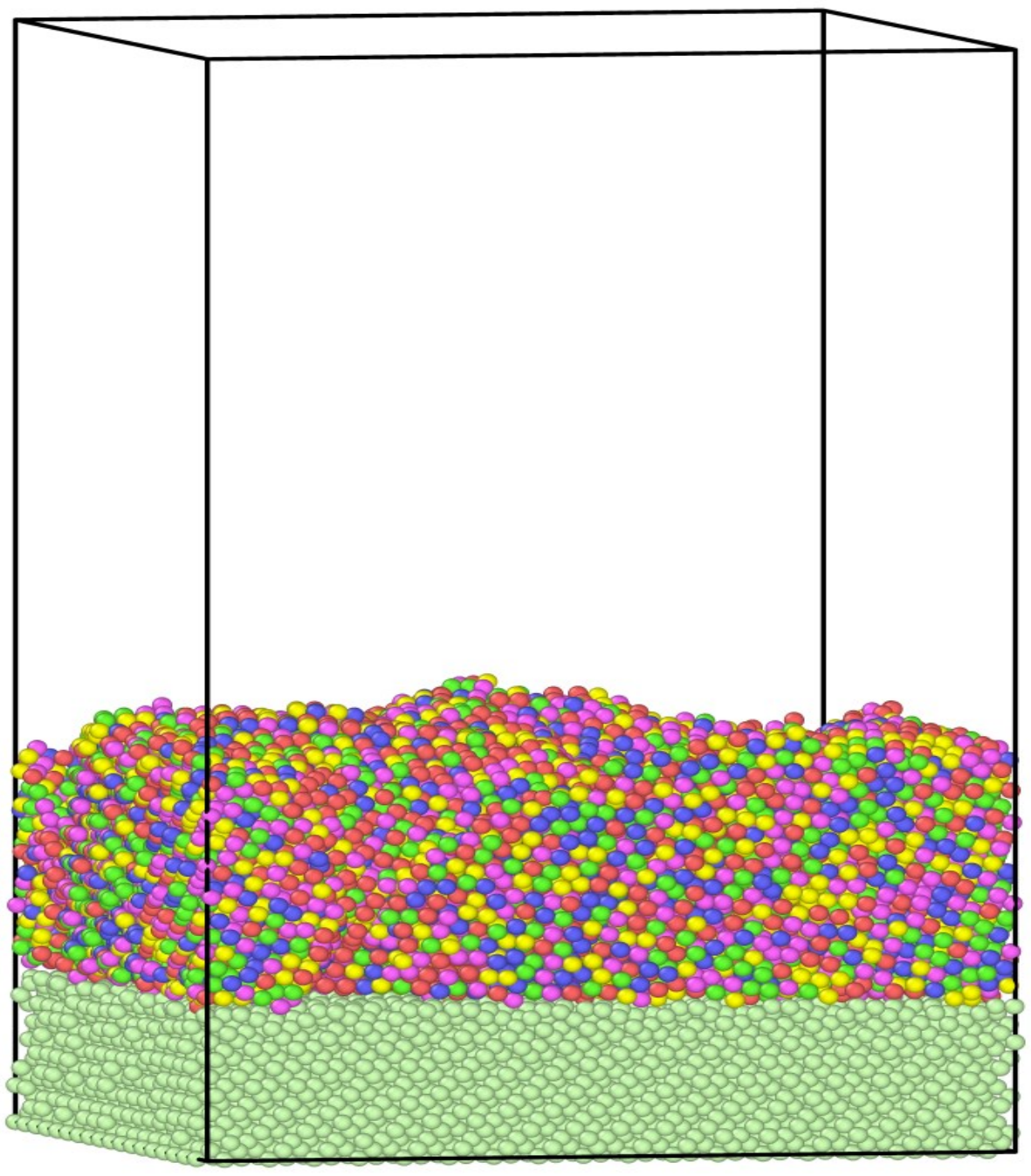

**Fig. 3** The AlCoCuFeNi thin film obtained after the 50 ns of simulation time

Following the results of common neighbor analysis (CNA), the process of crystallization started after ~ 5 ns of simulation, at the characteristic sizes of islands (clusters) of about 2 nm. At the end of the simulation, after the 50 ns of modeling, the simulated film contains a face-centered cubic (FCC) phase (content 20,5 %), a body-centered cubic (BCC) phase (content 13,2%), a hexagonal close-packed (HCP) phase (content 39,7%) (Fig.4) and an indefinite phase (content 26,6%), which, according to the analysis of the radial distribution of atoms (RDF), has an amorphous structure. It should be noted that in bulk AlCoCuFeNi alloys only the presence of FCC and BCC phases was recorded [15], but at the same time in melt-spun films, the formation of the nanotwin structure and stacking faults in the FCC phase was noted [16]. Thus, the specific features of the distribution of the HCP phase atoms (Fig.4) lead to the conclusion that this phase must be formed mainly of extrinsic (two HCP layers with an FCC layer between them) and intrinsic (two or more adjacent HCP layers) stacking faults in the lattice of the FCC phase. There are also twin boundaries between the FCC twins (a single layer composed of HCP atoms). As for the amorphous phase, its formation is explained by the high cooling rate that occurs during film deposition at which some of the atoms do not have time to rearrange and form a crystalline phase. In the context of the present simulation, this concerns mainly atomic layers located near the film surface, while crystallization processes take place in deeper layers.

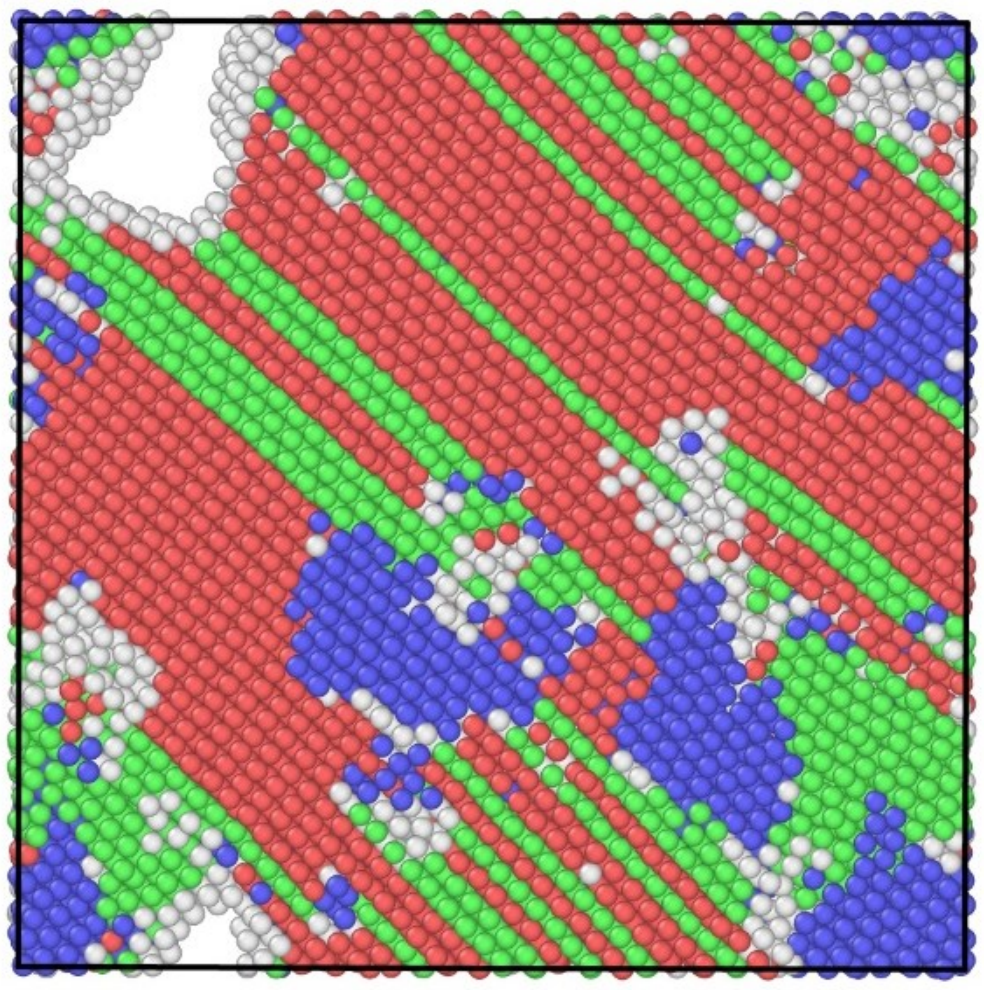

**Fig.4** Snapshot of AlCoCuFeNi thin film sliced along the substrate plane (top view). Atoms are colored according to CNA analysis results: ● – FCC, ● – BCC, ● – HCP, ● – amorphous phase

It is well known from the literature, that there are two main criteria by which the high-entropy alloys are usually characterized. This is the entropy of mixing $\Delta S_{mix}$ and the enthalpy of mixing $\Delta H_{mix}$. However, to predict the phase composition of HEAs, some additional parameters were proposed [1,2]. These parameters include in particular the valence electron concentration (VEC), the thermodynamic parameter Ω, which takes into account the melting temperature, mixing entropy, and the mixing enthalpy. The important parameter is an atomic-size difference between alloy components which is denoted as δ. Using the data from [17,18] we calculated $\Delta S_{mix}$, $\Delta H_{mix}$, δ, *VEC,* and *Ω* for the AlCoCuFeNi HEA (Tab.2).

**Table 2** Electronic, thermodynamic, and the atomic-size parameters of the AlCoCuFeNi HEA

| $\Delta S_{mix}$, J/(mol·K) | $\Delta H_{mix}$, kJ/mol | Ω | VEC | δ |
|---|---|---|---|---|
| 13.37 | -5.28 | 3.847 | 8.2 | 5.323 |

According to [1,2] the HEA alloys for which Ω≥1.1 and δ≤6.6 can form solid solutions without intermetallic compounds. However, simple (not ordered) solid solutions form if -15 kJ/mol < $\Delta H_{mix}$ < 5 kJ/mol. The other useful parameter is *VEC*, which has been proven suitable in determining the phase stability of high- entropy alloys. As pointed in [19] at $VEC \geq 8.0$, the sole FCC phase is expected in the alloy; at $6.87 \leq VEC < 8.0$, mixed FCC and BCC phases will co-exist and the sole BCC phase expected at $VEC < 6.87$. Thus, the simulation results, as well as the experimental results, generally confirm the validity of the above criteria, except for the VEC parameter. But, if the value of VEC is close to the boundary values, predictions of the phase compositions sometimes do not work [5].

Fig.5 presents the total RDF curves after 50 ns of simulation time. RDF patterns are obtained both for the film as a whole and for individual phases separated by CNA analysis.

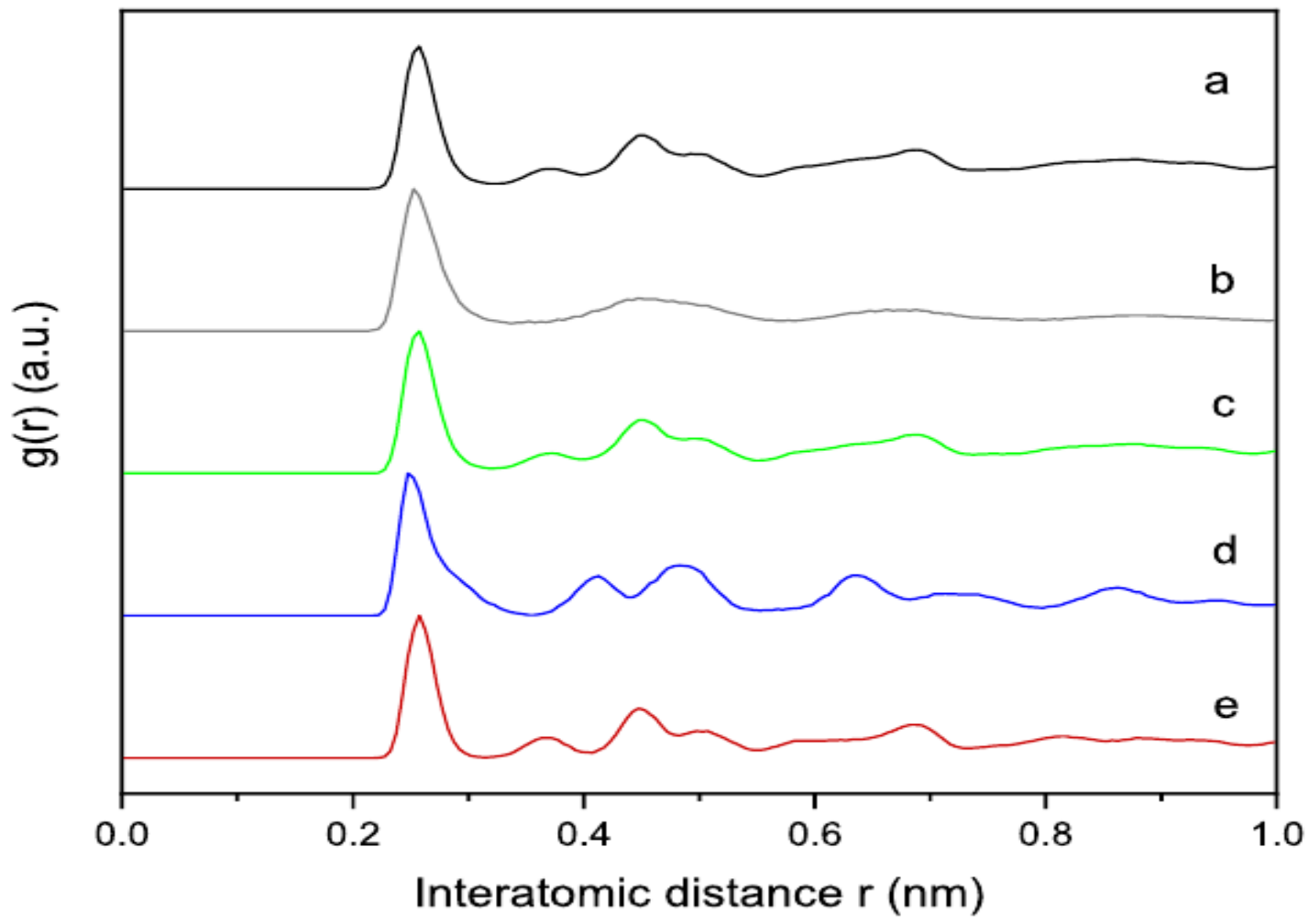

**Fig. 5** The RDF curves of AlCoCuFeNi thin film: a- whole film, b- amorphous phase, c- FCC phase, d- BCC phase, e- HCP phase

It is clearly visible, that the RDF form of an indefinite phase is intermediate between liquid and crystal, indicating an amorphous structure. The peaks on the RDF curves represent the successive distances between neighbors for different phases in the simulated film. So, from the RDF pattern for whole film these distances are: 1-st neighbor - 0.2575 nm, 2-st neighbor - 0.3725 nm, 3-st neighbor - 0.4525 nm, 4-st neighbor - 0.4925 nm.

Determining the distances between the nearest neighbors for the BCC and FCC lattices made it possible to estimate the lattice parameters for these phases. Tab. 3 shows the comparison between estimated parameters of the BCC and FCC lattices from the present MD simulation and the experimental results.

**Table 3** Calculated and experimental lattice parameters for AlCoCuFeNi HEA

| Lattice parameter *a,* nm | Experiment, bulk alloy | Experiment, melt-spun ribbon | Calculated |
|---|---|---|---|
| BCC phase | 0.2878 | 0.2872 | 0.2895 |
| FCC phase | 0.3624 | 0.3618 | 0.3670 |

As can be seen from Tab.3, calculated lattice parameters are in good agreement with the experimental results, taking into account that the estimated average film temperature after 50 ns simulation was above room temperature and was ~350 K.

## 4. Conclusions

Molecular dynamics simulation was performed to describe the processes of growth and crystallization of thin AlCoCuFeNi HEA films. The structural characteristics and phase composition of films were examined utilizing CNA and RDF. According to simulation results, the growth of AlCoCuFeNi films occurs via a formation of three-dimensional adatom clusters or islands with subsequent coarsening and coalescence. The estimated lattice parameters of the BCC and FCC phases are in good agreement with the lattice parameters of AlCoCuFeNi alloy obtained from the experiment.